\documentclass[12pt]{article}
\hoffset= - 2cm
\usepackage{amssymb,graphicx}
\usepackage{epstopdf}
\usepackage{amsmath,amsfonts}
\usepackage[normalem]{ulem}
\usepackage{epsfig}
\usepackage[dvipsnames]{xcolor}

\newcommand{\be}{\begin{equation}}
\newcommand{\ee}{\end{equation}}
\newcommand{\bea}{\begin{eqnarray}}
\newcommand{\eea}{\end{eqnarray}}

\usepackage{hyperref}
\definecolor{linkcolor}{HTML}{799B03}
\definecolor{urlcolor}{HTML}{799B03}
\hypersetup{pdfstartview=FitH,linkcolor=linkcolor,urlcolor=urlcolor,colorlinks=true}

\makeatletter
\newcommand*{\myfnsymbolsingle}[1]{%
  \ensuremath{%
    \ifcase#1% 0
    \or % 1
      *%
    \or % 2
      \dagger
    \or % 3
      \ddagger
    \or % 4
      1
    \or % 5
      2
    \or
      3
    \or
      4
    \or
      5
    \or
      6
    \or
      7
    \or
      8
    \else % >= 6
      \@ctrerr
    \fi
  }%
}
\makeatother

\usepackage{alphalph}
\newalphalph{\myfnsymbolmult}[mult]{\myfnsymbolsingle}{}
\renewcommand*{\thefootnote}{%
  \myfnsymbolmult{\value{footnote}}%
}

\begin{document}
\begin{flushright}
INR-TH-2020-030
\end{flushright}

\begin{center}
  {\LARGE \bf  Superluminality in beyond Horndeski theory
    with extra scalar field}

\vspace{10pt}

\vspace{20pt}
S. Mironov$^{a,c,d,e}$\footnote{sa.mironov\_1@physics.msu.ru},
V. Rubakov$^{a,b}$\footnote{rubakov@inr.ac.ru},
V. Volkova$^{a}$\footnote{volkova.viktoriya@physics.msu.ru}
\renewcommand*{\thefootnote}{\arabic{footnote}}
\vspace{15pt}

$^a$\textit{Institute for Nuclear Research of the Russian Academy of Sciences,\\
60th October Anniversary Prospect, 7a, 117312 Moscow, Russia}\\
\vspace{5pt}

$^b$\textit{Department of Particle Physics and Cosmology, Physics Faculty,\\
M.V. Lomonosov Moscow State University,\\
Vorobjevy Gory, 119991 Moscow, Russia}

$^c$\textit{Institute for Theoretical and Experimental Physics,\\
  Bolshaya Cheryomushkinskaya, 25, 117218 Moscow, Russia}

$^d$\textit{Moscow Institute of Physics and Technology,\\
Institutski pereulok, 9, 141701, Dolgoprudny, Russia}

$^e$\textit{Institute for Theoretical and Mathematical Physics,\\
M.V. Lomonosov Moscow State University, 119991 Moscow, Russia}
\end{center}

\vspace{5pt}

\begin{abstract} 
We study the superluminality issue in
    beyond Horndeski theory with additional 
    scalar field, which is minimally coupled
    to gravity and has no second derivatives
    in the Lagrangian.
  We present the quadratic action for perturbations 
  in  cosmological backgrounds, stability conditions and
  expressions for sound speeds. We find that in the case of
  conventional additional scalar whose flat-space propagation speed is that
  of light, one of the modes in interacting theory is
  necessarily superluminal when this scalar rolls, even
  arbitrarily slowly.
  {This result holds in any theory of
  the beyond Horndeski class (with 6 arbitrary functions in the
  Lagrangian) and for any stable rolling background.}
  More generally, 
    the requirement of the
  absence of superluminality imposes non-trivial constraints on the
  structure of the theory.

\end{abstract}

%%%%%%%%%%%%%%%%%%%%%%%%%%%%%%%%%%%%%%%%%%%%%%%%%%%%%%%%%%%%%%%%%%%%%%%%%
\section{Introduction}
\label{sec:intro}
{A class of scalar-tensor} theories of
{gravity --- Horndeski 
theories~\cite{Horndeski} and their 
extensions~\cite{Zuma,GLVP,KobaRev} --} has proved itself promising
candidate for supporting
various cosmological scenarios including those without the initial
{singularity.}
  What makes (beyond) Horndeski theories and
{more} general 
DHOST theories~\cite{DHOST} suitable for constructing 
non-singular cosmological solutions 
is their ability to violate the 
Null Energy 
   Condition (NEC)/Null Convergence Condition (NCC)
  while leaving
the stability of 
{the background} intact (for a review see, e.g.,
Ref.~\cite{RubakovNEC}).

{Even though the NEC/NCC can be safely violated in
    unextended Horndeski theories, the latter
do not enable one to construct non-singular spatially flat
  cosmological} solutions which are stable during
  the entire evolution~\cite{LMR,Koba_nogo}.
{On the contrary,}
beyond Horndeski and DHOST theories admit 
{completely} stable 
cosmologies 
with a bouncing or Genesis stage, 
see Refs.~\cite{Cai,CreminelliBH,RomaBounce,CaiBounce,genesisGR,chineseBounce2} 
for
{specific} examples and Refs.~\cite{KobaRev,Khalat} 
for topical reviews.

Another characteristic feature of   
{modified gravities}
is potential appearance of superluminal perturbations.
The issue of superluminality in Horndeski theories has been addressed 
from different viewpoints, see 
Refs.~\cite{BabVikMukh,gen_original,subl_gen,MatMat,Unbraiding} 
and references therein.
One of the most
{striking findings is that
at least in
a pure Horndeski Genesis model of Ref.~\cite{subl_gen},
addition of even
tiny amount of external
matter (ideal fluid) inevitably induces 
  superluminality in some otherwise healthy region of
  phase
  space}~\cite{MatMat}. The latter fact
{is troublesome
  (provided one would like to
    avoid superluminality altogether in view of arguments of
    Ref.~\cite{superlum1}),
  since nothing appears to prevent adding extra 
  fluid to Horndeski theory.}
  {Likewise,
superluminality has been shown to occur in other
stable non-singular cosmological backgrounds: in Cuscuton gravity~\cite{Quintin:2019orx}
and in DHOST theory~\cite{Ilyas:2020qja}.}

A step forward 
has been recently made in Ref.~\cite{sublum}, where
{a beyond Horndeski model admitting}
a completely stable
bouncing solution 
{has been analyzed from the
viewpoint of
potential superluminality.} As opposed to
{Genesis-supporting} unextended
Horndeski  
{model} with external matter~\cite{MatMat},
it {has been} shown that 
a specifically designed beyond Horndeski 
Lagrangian, which
{on its own admits  a stable} and subluminal 
bouncing
{solution, remains free of 
superluminalities} 
upon adding extra matter
{in the form of
perfect fluid with}  equation
of state
parameter $w \leq 1/3$ (or even somewhat larger).

{On the other hand, by}
analysing the general expressions for the sound
speeds {of} scalar modes in the system 
{``beyond Horndeski + perfect
  fluid'',
  it has been} found that for 
{$w$ equal or close to 1,} one of the scalar
propagation speeds
inevitably becomes superluminal. The latter statement holds 
irrespectively of the cosmological scenario one considers, and is true
for the most general beyond Horndeski
{theory~\cite{sublum}. This has to do with the fact,
already {noticed} 
in Refs.~\cite{GLVP,Gleyzes}, that} due to specific structure of 
beyond Horndeski Lagrangian,
{there is kinetic mixing between
  matter and Galileon perturbations, and hence
the sound speeds of 
  both scalar modes get modified (the superluminal  one is
predominantly
sound wave in matter). The results of
Ref.~\cite{sublum} imply} that in beyond Horndeski
{theory with}  an additional 
minimally coupled
conventional 
scalar field, {whose flat-space propagation speed is that
  of light,
  one of the scalar modes is superluminal
when this extra field has small but non-zero background kinetic energy.
The main purpose of this note is to derive this property explicitly.}
{We emphasize that superluminality is 
generic
  for beyond Horndeski theory (whose action is given by
  eq.~\eqref{eq:action_setup})
       {in the presence of 
         {additional} minimally coupled conventional
  scalar field;}
 this  property holds for any {choice of} 
 Lagrangian functions
  provided that at least one of the beyond Horndeski terms does
  not vanish.
  {This result} {applies to}
  a completely arbitrary {stable}
  cosmological background with rolling scalar (except for configurations of
  measure zero in the phase space),
  irrespectively of whether NEC/NCC
  is violated or not.}

In Sec.~\ref{sec:1} 
{we adopt the covariant
formulation and notations of Refs.~\cite{sublum,KobaRev} and derive} 
the quadratic action 
for perturbations about
a cosmological background in beyond Horndeski theory in
{the}
presence
of an additional minimally coupled scalar field of the most 
general type\footnote{{This generalizes the formulas
given in Ref.~\cite{KobaRev}; 
  similar results have been obtained} in ADM formalism in 
  Refs.~\cite{GLVP,Gleyzes}.}.
{In this way we}
obtain
stability conditions and
{prepare for the calculation of
the propagation speeds of perturbations in Sec.~\ref{sec:speeds}.}
Our expressions for speeds show explicitly that
{once the flat-space
speed of the scalar is equal to 1, one of the modes is superluminal
in ``beyond Horndeski + scalar field'' system provided the scalar
field background is rolling, even slowly.} 
We discuss the results in Sec.~4.

%%%%%%%%%%%%%%%%%%%%%%%%%%%%%%%%%%%%%%%%%%%%%%%%%%%%%%%%%%%%%%%%%%%%%%%%%
\section{Beyond Horndeski theory with additional scalar field}
\label{sec:1}

\subsection{{Setup}}
%%%%%%%%%%%%%%%%%%%%%%%%%%%%%%%%%%%%%%%%%%%%%%%%%%%%%%%%%%%%%%%%%%%%%%%%%
In this section we specify our setup and give
{background} equations in spatially flat FLRW geometry
{(our signature convention is mostly negative).}

{We} consider beyond Horndeski theory of the most general form:
\begin{subequations}
\label{eq:action_setup}
\begin{align}
S_{\pi}&=\int\mathrm{d}^4x\sqrt{-g}\left(\mathcal{L}_2 + \mathcal{L}_3 + \mathcal{L}_4 + \mathcal{L}_5 \right),\\
\mathcal{L}_2&=F(\pi,X),\\
\mathcal{L}_3&=K(\pi,X)\Box\pi,\\
\mathcal{L}_4&=-G_4(\pi,X)R+2G_{4X}(\pi,X)\left[\left(\Box\pi\right)^2-\pi_{;\mu\nu}\pi^{;\mu\nu}\right] \nonumber \\
% \hspace{5.5cm}
&+ F_4(\pi,X)\epsilon^{\mu\nu\rho}_{\quad\;\sigma}\epsilon^{\mu'\nu'\rho'\sigma}\pi_{,\mu}\pi_{,\mu'}\pi_{;\nu\nu'}\pi_{;\rho\rho'},\\
\mathcal{L}_5&=G_5(\pi,X)G^{\mu\nu}\pi_{;\mu\nu}+\frac{1}{3}G_{5X}\left[\left(\Box\pi\right)^3-3\Box\pi\pi_{;\mu\nu}\pi^{;\mu\nu}+2\pi_{;\mu\nu}\pi^{;\mu\rho}\pi_{;\rho}^{\;\;\nu}\right]
\nonumber \\
& 
% \hspace{5.5cm}
+F_5(\pi,X)\epsilon^{\mu\nu\rho\sigma}\epsilon^{\mu'\nu'\rho'\sigma'}\pi_{,\mu}\pi_{,\mu'}\pi_{;\nu\nu'}\pi_{;\rho\rho'}\pi_{;\sigma\sigma'},
\end{align}
\end{subequations}
where $\pi$ is a scalar field sometimes dubbed Galileon,
$X=g^{\mu\nu}\pi_{,\mu}\pi_{,\nu}$,
$\pi_{,\mu}=\partial_\mu\pi$,
$\pi_{;\mu\nu}=\nabla_\nu\nabla_\mu\pi$,
$\Box\pi = g^{\mu\nu}\nabla_\nu\nabla_\mu\pi$,
$G_{4X}=\partial G_4/\partial X$, etc. 
The functions $F$, $K$, $G_4$ and $G_5$
are characteristic of unextended Horndeski theories, while
non-vanishing $F_4$ and $F_5$ extend the theory to beyond Horndeski
type. Along with the scalar field of beyond Horndeski type we consider
another scalar field $\chi$ in the form of k-essence
\be
\label{eq:action_setup_kess}
S_{\chi} = \int\mathrm{d}^4x\sqrt{-g} \,P(\chi,Y), \quad Y = g^{\mu\nu}\chi_{,\mu}\chi_{,\nu}\,.
\ee
The Lagrangian in eq.~\eqref{eq:action_setup_kess} describes 
a minimally coupled
scalar field $\chi$ of the most general type
{(assuming the absence of second derivatives in the
  Lagrangian)}.

{In flat space-time and for spatially homogeneous background
  (possibly rolling, $Y=\dot{\chi}^2 \neq 0$), the stability conditions
  for the scalar field $\chi$ have standard form
  \be
  P_Y > 0 \; , \;\;\;\;\;  R \equiv P_Y +2 Y P_{YY} >0 \; ,
  \label{apr24-20-1}
  \ee
  while flat-space propagation speed of perturbations is
  \be
\label{eq:cm}
c_m^2 = \dfrac{P_Y}{R} \; .
\ee
Our main result on superluminality in
Sec.~\ref{sec:speeds} applies most straightforwardly to the conventional
scalar field with
\be
P= \dfrac{1}{2} Y - V(\chi) \; ,
\label{apr25-20-2}
\ee
but in this Section we proceed in full generality and do not make
any assumptions on the form of the function $P(\chi, Y)$.
}

In what follows we consider cosmological setting {with}
spatially flat 
FLRW {metric}
and homogeneous background scalar fields $\pi=\pi(t)$ and $\chi=\chi(t)$
{($t$ is cosmic time).}
Then the background {gravitational} equations following from the 
action $S_{\pi}+S_{\chi}$ read
\begin{subequations}
\label{eq:Einstein_kess}
\begin{align}
%\label{eq:dg00_kess}
\nonumber
  \delta g^{00}: \;\;
&F-2F_XX-6HK_XX\dot{\pi}+K_{\pi}X+6H^2G_4
+6HG_{4\pi}\dot{\pi}-24H^2X(G_{4X}+G_{4XX}X)
\\\nonumber&+12HG_{4\pi X}X\dot{\pi}
-2H^3X\dot{\pi}(5G_{5X}+2G_{5XX}X)+3H^2X(3G_{5\pi}+2G_{5\pi X}X)
\\&+6H^2X^2(5F_4+2F_{4X}X)
+6H^3X^2\dot{\pi}(7F_5+2F_{5X}X) + P - 2 P_Y Y = 0,
\label{eq:dg00_kess}\\
%\label{eq:dgii_kess}
\nonumber
\delta g^{ii}: \;\;
&F-X(2K_X\ddot{\pi}+K_\pi)+2(3H^2+2\dot{H})G_4-12H^2G_{4X}X
-8\dot{H}G_{4X}X-8HG_{4X}\ddot{\pi}\dot{\pi}
\\\nonumber&-16HG_{4XX}X\ddot{\pi}\dot{\pi}
+2(\ddot{\pi}+2H\dot{\pi})G_{4\pi}+4XG_{4\pi X}(\ddot{\pi}-2H\dot{\pi})+2XG_{4\pi\pi}
\\\nonumber&-2XG_{5X}(2H^3\dot{\pi}+2H\dot{H}\dot{\pi}+3H^2\ddot{\pi})+G_{5\pi}(3H^2X+2\dot{H}X+4H\ddot{\pi}\dot{\pi})-4H^2G_{5XX}X^2\ddot{\pi}
\\\nonumber&+2HG_{5\pi X}X(2\ddot{\pi}\dot{\pi}-HX)
+2HG_{5\pi\pi}X\dot{\pi}+2F_4X(3H^2X+2\dot{H}X+8H\ddot{\pi}\dot{\pi})
\\\nonumber&+8HF_{4X}X^2\ddot{\pi}\dot{\pi}+4HF_{4\pi}X^2\dot{\pi}+6HF_5X^2(2H^2\dot{\pi}+2\dot{H}\dot{\pi}+5H\ddot{\pi})
+12H^2F_{5X}X^3\ddot{\pi}
\\&+6H^2F_{5\pi}X^3 + P= 0,
\label{eq:dgii_kess}
\end{align}
\end{subequations} 
where
$P_Y \equiv \partial P/\partial Y$, and $H=\dot{a}/a$ is 
the Hubble parameter.
The field equation for the additional scalar field $\chi$ is:
\be
\label{eq:background_kess}
\ddot{\chi} + 3 c_m^2 H \dot{\chi} - \dfrac{P_{\chi} 
- 2 Y P_{\chi Y}}{2\,R} = 0 \; .
\ee
{The} field equation for Galileon $\pi$ follows from 
{the gravitational}
equations~\eqref{eq:Einstein_kess}, their derivatives 
and eq.~\eqref{eq:background_kess}, so we do not give it here for brevity.

%%%%%%%%%%%%%%%%%%%%%%%%%%%%%%%%%%%%%%%%%%%%%%%%%%%%%%%%%%%%%%%%%%%%%%%%%
\subsection{Quadratic action and stability conditions}
\label{sec:stability}

{To} address stability {and superluminality} issues,
{we} calculate the quadratic action {for
  perturbations about
homogeneous background} in terms {of}
propagating degrees of freedom (DOFs).
{We} make use of the standard ADM 
parametrization of the metric perturbations,
\begin{equation}
\label{eq:FLRW_perturbed}
\mathrm{d}s^2 = N^2 \mathrm{d}t^2 - \gamma_{ij}(\mathrm{d}x^i+ N^i \mathrm{d}t)(\mathrm{d}x^j+N^j \mathrm{d}t),
\end{equation}
where 
\be
\label{eq:ADM}
N = 1+\alpha, \qquad N_i = \partial_i\beta, \qquad
\gamma_{ij}= a^2(t) e^{2\zeta} \left(\delta_{ij} + h_{ij}^T +
\dfrac12 h_{ik}^T {h^{k\:T}_j}\right),
\ee
{and we have already used some part of gauge 
  freedom by {setting the longitudinal part of
  $\delta \gamma_{ij}$ equal to zero,} $\partial_i\partial_j E = 0$.}
Here the scalar sector consists of $\alpha$, $\beta$, $\zeta$ 
from eq.~\eqref{eq:FLRW_perturbed} and scalar field perturbations
$\delta\pi$ and
\[
{\delta\chi \equiv \omega\; ,}
\]
while $h_{ij}^T$ 
denote tensor modes 
($h_{ii}^T = 0, \partial_i h_{ij}^T = 0$).
Like in Ref.~\cite{sublum} we adopt the unitary gauge {where}
$\delta\pi = 0$.
{Then} the quadratic action for beyond Horndeski 
theory~\eqref{eq:action_setup} reads~\cite{RomaBounce}:
\begin{equation}
\label{eq:pert_action_setup}
\begin{aligned}
S^{(2)}_{\pi}=\int\mathrm{d}t\,\mathrm{d}^3x \,a^3\Bigg[\left(\dfrac{\mathcal{{G}_T}}{8}\left(\dot{h}^T_{ik}\right)^2-\dfrac{\mathcal{F_T}}{8a^2}\left(\partial_i h_{kl}^T\right)^2\right)+
\left(-3\mathcal{{G}_T}\dot{\zeta}^2+\mathcal{F_T}\dfrac{(\nabla\zeta)^2}{a^2}+\Sigma\alpha^2
\right.\\\left.
-2(\mathcal{G_T}+\mathcal{D}\dot{\pi})\alpha\dfrac{\nabla^2\zeta}{a^2}+6\Theta\alpha\dot{\zeta}-2\Theta\alpha\dfrac{\nabla^2\beta}{a^2}
+2\mathcal{{G}_T}\dot{\zeta}\dfrac{\nabla^2\beta}{a^2}\right)\Bigg],
\end{aligned}
\end{equation}
with $(\nabla\zeta)^2 = \delta^{ij} \partial_i \zeta \partial_j \zeta$,
$\nabla^2 = \delta^{ij} \partial_i \partial_j$
and
% \begin{eqnarray}
\begin{subequations}
\begin{align}
\label{eq:GT_coeff_setup}
&\mathcal{G_T}=2G_4-4G_{4X}X+G_{5\pi}X-2HG_{5X}X\dot{\pi} + 2F_4X^2+6HF_5X^2\dot{\pi},
\\
&\mathcal{F_T}=2G_4-2G_{5X}X\ddot{\pi}-G_{5\pi}X,\\
\label{eq:D_coeff_setup}
&\mathcal{D}=-2F_4X\dot{\pi}-6HF_5X^2,\\
%\label{eq:Theta_coeff_setup}
&\Theta=-K_XX\dot{\pi}+2G_4H-8HG_{4X}X-8HG_{4XX}X^2+G_{4\pi}\dot{\pi}+2G_{4\pi X}X\dot{\pi}-5H^2G_{5X}X\dot{\pi}\nonumber\\
&-2H^2G_{5XX}X^2\dot{\pi}+3HG_{5\pi}X+2HG_{5\pi X}X^2
% \\\nonumber&
+10HF_4X^2+4HF_{4X}X^3+21H^2F_5X^2\dot{\pi}
 \nonumber\\
 %\nonumber
 &
 +6H^2F_{5X}X^3\dot{\pi},
 \label{eq:Theta_coeff_setup}
\\
%\end{align}
% \end{eqnarray}
%\begin{subequations}
%\bea
%\begin{aligned}
%\label{eq:Sigma_coeff_setup}
&\Sigma=F_XX+2F_{XX}X^2+12HK_XX\dot{\pi}+6HK_{XX}X^2\dot{\pi}-K_{\pi}X-K_{\pi X}X^2-6H^2G_4
\nonumber\\
% \nonumber
&+42H^2G_{4X}X+96H^2G_{4XX}X^2+24H^2G_{4XXX}X^3-6HG_{4\pi}\dot{\pi}-30HG_{4\pi X}X\dot{\pi}
\nonumber\\
 \nonumber
&-12HG_{4\pi XX}X^2\dot{\pi}+30H^3G_{5X}X\dot{\pi}+26H^3G_{5XX}X^2\dot{\pi}+4H^3G_{5XXX}X^3\dot{\pi}-18H^2G_{5\pi}X\\
 \nonumber
&-27H^2G_{5\pi X}X^2-6H^2G_{5\pi XX}X^3-90H^2F_4X^2-78H^2F_{4X}X^3-12H^2F_{4XX}X^4\\
% \nonumber
&-168H^3F_5X^2\dot{\pi}-102H^3F_{5X}X^3\dot{\pi}-12H^3F_{5XX}X^4\dot{\pi}.
\label{eq:Sigma_coeff_setup}
%\end{aligned}
%\eea
\end{align}
\end{subequations}
{The first round brackets in eq.~\eqref{eq:pert_action_setup} 
describe tensor sector, while the second ones refer to scalar modes.}
{The} quadratic action for 
k-essence~\eqref{eq:action_setup_kess} is as follows:
\be
\begin{aligned}
\label{eq:quadr_action_kess}
S^{(2)}_{\chi} = \int \mathrm{d}t\,\mathrm{d}^3x \,a^3 \left[ 
Y R \,\alpha^2 - 2 \dot{\chi} R \, \alpha\dot{\omega} 
+ 2\dot{\chi} P_Y \, \omega\dfrac{\nabla^2\beta}{a^2}
+ R\, \dot{\omega}^2 - P_Y\, \dfrac{(\nabla\omega)^2}{a^2} 
\right.\\ \left. 
- 6 \dot{\chi} P_Y\, \dot{\zeta}\omega 
+ (P_{\chi}-2 Y P_{\chi Y})\,\alpha\omega
+\Omega \, \omega^2
\right],
\end{aligned}
\ee
where  
$\Omega = P_{\chi\chi}/2 - 3 H \dot{\chi} P_{\chi Y} - Y P_{\chi\chi Y}  - \ddot{\chi} (P_{\chi Y} + 2 Y P_{\chi YY})$. 
When deriving the actions~\eqref{eq:pert_action_setup} 
and~\eqref{eq:quadr_action_kess} 
we used background equations~\eqref{eq:Einstein_kess}, which 
made the {terms with} $\alpha\zeta$, $\zeta^2$ and 
$\zeta\omega$ vanish. 

Let us for a moment concentrate on the scalar sector.
According to the form of actions~\eqref{eq:pert_action_setup} 
and~\eqref{eq:quadr_action_kess}, $\alpha$ and $\beta$ are 
non-dynamical variables, so varying $S^{(2)}_{\pi} +S^{(2)}_{\chi}$ 
with respect to $\alpha$ and $\beta$ gives the following 
constraint equations, respectively:
\begin{subequations}
\label{eq:constraints}
\begin{align}
\label{eq:beta}
& 
% {\delta g^{0}_{0}}: \quad 
\Sigma \alpha - \left(\mathcal{G_T} + \mathcal{D} \dot{\pi}\right) 
\dfrac{(\nabla^2\zeta)}{a^2} +3 \Theta \dot{\zeta} - 
\Theta  \dfrac{(\nabla^2\beta)}{a^2}
% \\&\nonumber
+ Y R\,\alpha -  \dot{\chi} R\, \dot{\omega} 
+  \frac12 (P_{\chi} - 2 Y P_{\chi Y})\,\omega   = 0,\\
\label{eq:alpha}
&  
% {\delta g^{0}_{i}}: \quad 
\hspace{7cm}\Theta \alpha -\mathcal{{G}_T} \dot{\zeta} - \dot{\chi}P_Y \,\omega =0.
\end{align}
\end{subequations}
{By} solving eqs.~\eqref{eq:beta} and~\eqref{eq:alpha} for 
$(\nabla^2\beta)/{a^2}$ and $\alpha$ and substituting the result
{back} into
actions~\eqref{eq:pert_action_setup} and~\eqref{eq:quadr_action_kess}, 
one arrives
{at} the quadratic
action for scalar DOFs 
in terms of dynamical curvature perturbation $\zeta$ and
scalar field perturbation $\omega$:
\be
\label{eq:quadratic_action_final}
S^{(2)}_{\pi+\chi} = \int \mathrm{d}t\,\mathrm{d}^3x \,a^3
\left[G_{AB} \dot{v}^A \dot{v}^B 
- \dfrac{1}{a^2} F_{AB} \nabla_i\,{v^A} \nabla^i\,{v^B}+ 
\Psi_1\dot{\zeta}\omega + \Psi_2 \omega^2\right],
\ee
where $A,B = 1,2$ and $v^1 = \zeta$, $v^2 = \omega$. Even though 
coefficients $\Psi_1$ and $\Psi_2$  
{are} irrelevant for {kinetic
  stability (absence of ghosts and gradient instabilities) as well as
propagation} 
speeds of $\zeta$ and $\omega$, they
are given in Appendix for completeness. 
Kinetic matrices $G_{AB}$ and $F_{AB}$
have the following {forms}:
\be
G_{AB} = 
\begin{pmatrix}
\mathcal{{G}_S} + \dfrac{\mathcal{{G}_T}^2}{\Theta^2} YR 
& -\dfrac{\mathcal{{G}_T}}{\Theta} \dot{\chi}R \\
-\dfrac{\mathcal{{G}_T}}{\Theta} \dot{\chi}R 
& R
\end{pmatrix},
\;\;
F_{AB} = 
\begin{pmatrix}
\mathcal{{F}_S} 
& -\dfrac{\left(\mathcal{G_T} + \mathcal{D} \dot{\pi}\right)}{\Theta}
\dot{\chi} P_Y\\
-\dfrac{\left(\mathcal{G_T} + \mathcal{D} \dot{\pi}\right)}{\Theta}
\dot{\chi} P_Y & 
 P_Y
\end{pmatrix}\; ,
\ee
  {where
\begin{subequations}
  \begin{align}
  \mathcal{G_S} & = \dfrac{\Sigma\mathcal{{G}_T}^2}{\Theta^2}+3\mathcal{{G}_T},
  \\
  \mathcal{{F}_S} &= \dfrac{1}{a}\dfrac{\mathrm{d}}{\mathrm{d}t}
  \left[ \dfrac{a \;\mathcal{{G}_T}\left(\mathcal{G_T} + \mathcal{D} \dot{\pi}\right)}{\Theta}\right]
  -\mathcal{F_T}\; .
  \end{align}
\label{may21-20-1}
\end{subequations}}
It is worth noting that  both $\mathcal{G_S}$ and $\mathcal{{F}_S}$ are
  {generally}
singular at $\Theta =0$ ($\Theta$-crossing, or $\gamma$-crossing in
terminology of Refs.~\cite{Ijjas:2016tpn,Ijjas:2017pei}). However,
no singularity exists at  $\Theta =0$ in the Newtonian
gauge~\cite{Ijjas:2017pei}, and the perturbations
are non-singular
in the unitary gauge {as well}~\cite{Mironov:2018oec}. 
Thus, the system is
well behaved at the moment of time
when $\Theta=0$.

Now we can formulate the stability conditions for beyond Horndeski
theories with additional scalar field in {the}
cosmological setting.
Recalling the tensor part of quadratic action 
in eq.~\eqref{eq:pert_action_setup}, we see that the tensor sector is 
free of ghosts and gradient instabilities provided {that}
\be
\label{eq:stability_conditions_tensor}
\mathcal{G_T}>0, \quad \mathcal{F_T}>0.
\ee
Let us note here that stability 
conditions~\eqref{eq:stability_conditions_tensor} 
have retained their form as compared to the case of pure 
beyond Horndeski, see e.g. Ref.~\cite{Khalat}. 
However, since generally the coefficient 
$\mathcal{G_T}$~\eqref{eq:GT_coeff_setup} involves the Hubble parameter,
the stability of gravitational waves gets affected by the additional
k-essence through the Friedmann 
equation~\eqref{eq:dg00_kess}.

As for the scalar modes, it follows from 
action~\eqref{eq:quadratic_action_final} that scalar sector is 
{free of ghosts and gradient instabilities
iff} both kinetic matrices are positive definite 
{($G_{11}, G_{22}>0$, $\det G > 0$
  and $F_{11}, F_{22}>0$, $\det F > 0 $):
\be
\label{eq:stability_conditions_scalar}
\mathcal{{G}_S} > 0 \; , 
\quad \mathcal{{F}_S}>0,  \quad R > 0\; ,
\quad  P_Y>0 \;, \quad \mathcal{{F}_S} - Y P_Y\dfrac{\left(\mathcal{G_T} + \mathcal{D} \dot{\pi}\right)^2}{\Theta^2} > 0.
\ee
The first four conditions are formally the same as the stability
conditions in pure beyond Horndeski theory and pure k-essence theory
(extra scalar field affects $\mathcal{{G}_S}$ and 
$\mathcal{{F}_S}$ through the Hubble parameter only)
while the last condition is specific to the interacting theory.}

%%%%%%%%%%%%%%%%%%%%%%%%%%%%%%%%%%%%%%%%%%%%%%%%%%%%%%%%%%%%%%%%%%%%%%%%%
\section{Superluminality {due to conventional}
  scalar field}
\label{sec:speeds}

Let us now turn to the propagation speeds of {perturbations}.
The sound speed squared for tensor perturbations follows immediately 
from action~\eqref{eq:pert_action_setup}:
\be
\label{eq:tensor_speed}
c_{\mathcal{T}}^2 = \dfrac{\mathcal{F_T}}{\mathcal{G_T}}.
\ee
Again, $c_{\mathcal{T}}^2$ has a standard form, but in fact
the tensor sound speed changes upon 
introducing additional k-essence due to
new contributions in eq.~\eqref{eq:dg00_kess} and, hence,
the modified Hubble parameter. 

In the scalar sector,  
the propagation speeds of $\zeta$ and $\omega$ are given by eigenvalues
of matrix $G_{AB}^{-1}F_{AB}$:
\be
  \label{eq:matrix_final}
G_{AB}^{-1} F_{AB} = 
\begin{pmatrix}
\dfrac{\mathcal{F_S}}{\mathcal{{G}_S}} - \dfrac{\left(\mathcal{G_T} + \mathcal{D} \dot{\pi}\right)\mathcal{{G}_T}}{\Theta^2} 
\dfrac{Y P_Y}{\mathcal{{G}_S}} 
&
-\dfrac{\dot{\chi} P_Y}{\mathcal{{G}_S}} \dfrac{\mathcal{D}\dot{\pi}}{\Theta}
\\
\dfrac{\mathcal{{G}_T}}{\Theta}\dot{\chi}\left[ 
\dfrac{\mathcal{F_S}}{\mathcal{{G}_S}} - \dfrac{\left(\mathcal{G_T} + \mathcal{D} \dot{\pi}\right)\mathcal{{G}_T}}{\Theta^2} 
\dfrac{Y P_Y}{\mathcal{{G}_S}} \right] - \dfrac{\left(\mathcal{G_T} + \mathcal{D} \dot{\pi}\right)}{\Theta}\dfrac{\dot{\chi} P_Y}{R} \;\;\;
&
c_m^2- \dfrac{Y P_Y}{\mathcal{{G}_S}} \dfrac{\mathcal{{G}_T}\left(\mathcal{D}\dot{\pi}\right)}{\Theta^2}
\end{pmatrix} \; .
\ee
%%%%%%
{Explicitly, the speeds are} (recall that $c_m^2 = P_Y/R$):
\bea
\label{eq:speeds_eigenvalues}
{c_{\mathcal{S}\, \pm}^2} &=& \dfrac12 c_m^2+
\dfrac{1}{2} 
\left[ 
\dfrac{\mathcal{F_S}}{\mathcal{{G}_S}} - \dfrac{Y P_Y}{\mathcal{{G}_S}} 
\dfrac{\mathcal{{G}_T}(\mathcal{{G}_T}+2\mathcal{D}\dot{\pi})}{\Theta^2} 
% \phantom{\qquad\qquad\qquad\qquad}
\right.\\\nonumber
&&\left.\pm 
\sqrt{ \left(\dfrac{\mathcal{F_S}}{\mathcal{{G}_S}} - \dfrac{Y P_Y}{\mathcal{{G}_S}} \dfrac{\mathcal{{G}_T}(\mathcal{{G}_T}+2\mathcal{D}\dot{\pi})}{\Theta^2} + c_m^2
\right)^2 - 4 \, c_m^2 \left(\dfrac{\mathcal{F_S}}{\mathcal{{G}_S}} - 
\dfrac{Y P_Y}{\mathcal{{G}_S}} \dfrac{\left(\mathcal{{G}_T}+\mathcal{D}\dot{\pi}\right)^2}{\Theta^2} \right) } \,\,
\right].
\eea
{In accordance with the above remark, there is
 no singularity in the sound speeds
at $\Theta=0$. Indeed, 
  the speeds are finite
  as $\Theta \to 0$: one finds from eq.~\eqref{may21-20-1} that both
  ${\mathcal{F_S}}/{\mathcal{{G}_S}}$ and $\Theta^2 \,{\mathcal{{G}_S}}$
  are finite in this limit.}
{On the other hand, depending on the model, one of
  the sound speeds may become arbitrarily large in some region of parameter
  space, say, where ${\mathcal{{G}_S}} \to 0$ and
   ${\mathcal{{F}_S}}$ remains finite, cf. Ref.~\cite{MatMat}.}

{Now we  see} a considerable difference between the 
unextended Horndeski {and} beyond Horndeski theories. 
In the {unextended Horndeski} case, 
the coefficient 
$\mathcal{D}$ {vanishes} (see eq.~\eqref{eq:D_coeff_setup}),
so the matrix~\eqref{eq:matrix_final} {is} triangular and
the speed of perturbations in k-essence recovers its standard value 
$c_m^2$, 
while the propagation speed of Galileon perturbations is modified. 
{Indeed, for $\mathcal{D}=0$,}
eqs.~\eqref{eq:speeds_eigenvalues} reduce {to }
\be
\label{eq:speed_Horndeski}
{c_{\mathcal{S} \, -}^2}|_{\mathcal{D}=0}
= \dfrac{\mathcal{F_S}}{\mathcal{{G}_S}} -
\dfrac{Y P_Y}{\mathcal{{G}_S}} \dfrac{\mathcal{{G}_T}^2}{\Theta^2}, 
\quad 
{c_{\mathcal{S}\, +}^2}|_{\mathcal{D}=0} = c_m^2,
\ee
and we restore the results for Horndeski theory with k-essence $P(Y)$ 
given in Ref.~\cite{KobaRev}. On the contrary,
with $\mathcal{D} \neq 0$, {there
  is kinetic mixing between the
  scalars $\zeta$ and $\omega$, so both scalar speeds get modified,
in general agreement with Refs.~\cite{GLVP,Gleyzes}.} 

{The key observation is that eq.~\eqref{eq:speeds_eigenvalues}
  has the following form (cf. Ref.~\cite{sublum}):
  \be
c_{\mathcal{S}\, \pm}^2 = \dfrac12 (c_m^2 + \mathcal{A})
\pm  \dfrac12 \sqrt{(c_m^2 - \mathcal{A})^2 + \mathcal{B}},
\label{apr25-20-5}
\ee
where
\be
\mathcal{A}= \frac{\mathcal{F_S}}{\mathcal{G_S}} -
\frac{YP_Y}{\mathcal{G_S}} \, \frac{\mathcal{G_T}(\mathcal{G_T} + 2 \mathcal{D}\dot{\pi})}{ \Theta^2} \; , \;\;\;\;\;\;
\mathcal{B}= 4c_m^2\frac{YP_Y}{\mathcal{G_S}}
\frac{(\mathcal{D} \dot{\pi})^2}{ \Theta^2} \; .
  \nonumber
  \ee
  In stable and rolling background ($\mathcal{G_S}, P_Y >0$, $Y>0$),
    the coefficient   $\mathcal{B}$ is positive ($\mathcal{D} \neq 0$
    unless 
    the value of $Y$ and, hence, the Hubble parameter is fine-tuned, see
    eq.~\eqref{eq:D_coeff_setup}). This gives immediately
    \be 
    c_{\mathcal{S}\, +}^2 > c_m^2 \;\;\;\;\mbox{for}\;\; Y\neq 0 \; .
    \label{apr25-20-1}
    \ee
    So, if the flat-space propagation of the scalar perturbation
    $\omega$
    is luminal, $c_m=1$, then
    it becomes superluminal in the ``beyond Horndeski + scalar field''
    system. Equations  \eqref{eq:speeds_eigenvalues},
    \eqref{apr25-20-5} and \eqref{apr25-20-1} are
    our main results.
}

{\section{Discussion}}
\label{sec:conclusion-bis}

The interpretation of the result~\eqref{apr25-20-1} is most
straightforward in the case of the conventional scalar field $\chi$ 
with the Lagrangian~\eqref{apr25-20-2}.  
In that case one has $c_m = 1$ for any $Y$,
 and even tiny kinetic energy of rolling scalar background $\chi(t)$
 immediately yields superluminal propagation of one of the modes.
  {It is suggested (see, e.g., Ref.~\cite{deRham:2019ctd})
 that a covariant theory which is
fundamentally Lorentz invariant should recover a sound speed equal to unity
in the far UV limit ($k \rightarrow \infty$, where $k$ is spatial momentum),
even though for smaller $k$
perturbation modes
could be superluminal. Our result is independent of
$k$, so this is not  the case in theories we consider:
superluminality would occur even
as $k$ tends to infinity.
Therefore, if we decide to insist on Lorentz invariance of an
underlying theory and hence}
to avoid
superluminality for good, we have to conclude that  
in scalar-tensor theories with multiple scalar fields, none 
of these fields can be conventional and minimally coupled, 
as long as at least one of the scalar fields is of beyond Horndeski type.

More generally, if we insist on the absence of
  superluminality, the result \eqref{eq:speeds_eigenvalues},
  \eqref{apr25-20-5} implies a non-trivial constraint on the structure
  of ``beyond Horndeski + minimal quintessence'' systems:
  it is required that $c_{\mathcal{S}\, +} \leq 1$ everywhere in the part
  of the phase space $(\pi, \dot{\pi}, \chi, \dot{\chi})$
  where stability conditions \eqref{eq:stability_conditions_scalar}
  are satisfied.   
  In particular, this constraint forbids luminal flat-space propagation,
  $c_m=1$ (and, by continuity, $c_m$ close to 1),
  in any rolling background $Y\neq 0$, unless such a background
  is unstable for any $\pi$ and $\dot{\pi}$. Viewed differently,
  the constraint
that  $c_{\mathcal{S}\, +} \leq 1$  in the entire
``stable'' part of phase space suggests intricate
properties of the UV completion of the scalar-tensor theories considered in
this note, if such a UV completion exists and is Lorentz-invariant.

{We conclude by adding that it is certainly of interest to
  study the superluminality issue in more general DHOST theories
  coupled to conventional or
  $k$-essence scalar field(s), and also address phenomenological
  implications of our result, especially in models
for dark energy in the late-time Universe.}

%%%%%%%%%%%%%%%%%%%%%%%%%%%%%%%%%%%%%%%%%%%%%%%%%%%%%%%%%%%%%%%%%%%%%%%%%
\section*{Acknowledgements}
{We are indebted to an anonymous referee for valuable
  comments.}
  This work has been supported by Russian Science Foundation grant 19-12-00393.

%%%%%%%%%%%%%%%%%%%%%%%%%%%%%%%%%%%%%%%%%%%%%%%%%%%%%%%%%%%%%%%%%%%%%%%%%
\section*{Appendix}
In this Appendix we give explicit expressions {for}
coefficients 
$\Psi_1$ and $\Psi_2$ involved in the quadratic action~\eqref{eq:quadratic_action_final} for beyond Horndeski + k-essence $P(\chi,Y)$ 
{theory}:
\bea
&\Psi_1 = \dfrac{\mathcal{G_T}}{\Theta^2}\left[ 2 \dot{\chi} P_Y  
(\Sigma + Y R) + \Theta(P_{\chi} - 2 Y P_{\chi Y})
\right] , \\
&\Psi_2 = \Omega + \dfrac{\dot{\chi} P_Y}{\Theta} (P_{\chi} - 2 Y P_{\chi Y}) + \dfrac{Y P_Y^2}{\Theta^2}(\Sigma + Y R) + \dfrac{d}{dt}\Big[2Y P_Y\, R\Big],
\eea 
where
$$
\Omega = P_{\chi\chi}/2 - 3 H \dot{\chi} P_{\chi Y} - Y P_{\chi\chi Y}  - \ddot{\chi} (P_{\chi Y} + 2 Y P_{\chi YY}).
$$

%%%%%%%%%%%%%%%%%%%%%%%%%%%%%%%%%%%%%%%%%%%%%%%%%%%%%%%%%%%%%%%%%%%%%%%%%

\end{document}